\documentclass[prb,twocolumn,showpacs,superscriptaddress]{revtex4}
\usepackage{graphicx,amsmath,amssymb,natbib}

\newcommand{\eq}[1]{Eq.(\ref{#1})}
\newcommand{\fig}[1]{Fig.~\ref{#1}}

\newcommand{\Eq}[1]{Eq.(\ref{#1})}
\newcommand{\Fig}[1]{Fig.~\ref{#1}}

\newcommand{\kb}{ k_{\rm B} }

\newcommand{\gin}{ \gamma_{\rm IN} }

\newcommand{\lx}{ L_{\rm x} }
\newcommand{\ly}{ L_{\rm y} }
\newcommand{\lz}{ L_{\rm z} }
\newcommand{\D}{{\rm d}}

\begin{document}

\title{The isotropic-nematic interface in suspensions of hard rods:\\
  Mean-field properties and capillary waves.}

\author{S. Wolfsheimer}
\affiliation{Institut f\"ur Physik, Johannes Gutenberg-Universit\"at,
D-55099 Mainz, Staudinger Weg 7, Germany}

\author{C. Tanase}
\affiliation{Institute for Theoretical Physics, Utrecht University,
Leuvenlaan 4 3508 TD Utrecht, The Netherlands}

\author{K. Shundyak}
\affiliation{Instituut-Lorentz for Theoretical Physics, Niels
Bohrweg 2, Leiden, NL-2333 CA, The Netherlands}

\author{R. van Roij}
\affiliation{Institute for Theoretical Physics, Utrecht University,
Leuvenlaan 4 3508 TD Utrecht, The Netherlands}

\author{T. Schilling}
\affiliation{Institut f\"ur Physik, Johannes Gutenberg-Universit\"at,
D-55099 Mainz, Staudinger Weg 7, Germany}

\date{\today}

\begin{abstract} 
We present a study of the isotropic-nematic interface in
a system of hard spherocylinders. First we compare results from Monte Carlo
simulations and Onsager density functional theory for the interfacial profiles
of the orientational order parameter and the density. Those interfacial
properties that are not affected by capillary waves are in good agreement, 
despite the fact that Onsager theory overestimates the coexistence densities.
Then we show results of a Monte Carlo study of the capillary 
waves of the interface. In agreement with recent theoretical investigations
(Eur.~Phys.~J.~E {\bf 18} 407 (2005)) we find a strongly anistropic 
capillary wave spectrum. For the wave-numbers accessed in our 
simulations, the spectrum is 
quadratic, i.~e.~elasticity does not play a role. 
We conjecture that this effect is due to the strong bending rigidity of 
the director field in suspensions of spherocylinders.
\end{abstract}



\maketitle

\section{Introduction}

Anisotropic particles form liquid crystals at low enough temperatures
or high enough pressures.
Elongated particles, in particular, undergo a phase
transition from a phase in which particle orientations and
positions are disordered (isotropic phase) to a phase in which
orientations are aligned and positions are disordered (nematic
phase). In many materials the direction of preferred
alignment of the particles (the director
${\mathbf n}$) can be easily manipulated with electric or magnetic fields,  
which means that the optical properties of these materials can be tuned.
This has made liquid crystals the basis for a large range of
technological devices \cite{deGennes.Prost:1993}. 

Liquid crystals also pose fundamental
questions about the role which particle anisotropy plays in phase
transitions and  interfacial properties. 
In the 1940s  Onsager showed that the transition between the
isotropic phase (I) and the nematic phase (N) is of entropic nature
and that it can be explained by a simple
geometrical argument\cite{onsager:1949}. However, the calculation of
interfacial properties in the Onsager model proved to be far
more difficult than the prediction of the bulk phase transition. 

\subsection{Recent work on mean field properties of the IN interface}

Recently there have been several theoretical investigations
on the IN-interface 
\cite{mcmullen:1988, chen.noolandi:1992,chen:1993, koch.harlen:1999, schoot:1999,
velasco.mederos:2002, allen:2000*b, shundyak.roij:2001,
shundyak.roij:2003}. 
In the framework  of density functional theory (DFT)
all these studies showed that the interfacial tension of hard rods is minimal 
in the case of  parallel anchoring, i.~e.~when the nematic director is 
parallel to the plane of the interface.
Also, alignment is induced in the isotropic phase close to the interface. 
McMullen \cite{mcmullen:1988} showed by means of a variational method based 
on a square gradient expansion for the excluded volume that 
the density profile is monotonic in the case of parallel alignment.
This result was confirmed by the full numerical studies of Chen and
 Noolandi \cite{chen.noolandi:1992, chen:1993}. 
They also found small biaxial effects at the interface, and a value of
$\gin LD/k_BT=0.187\pm0.001$ for the interfacial tension $\gin$, 
where $L$ is the length of the particle and $D$ its diameter.
This value was $50\%$ lower than the result of the variational study by McMullen.
Shundyak and van Roij \cite{shundyak.roij:2001} 
developed a numerical method for a full description of the excluded volume 
based on the assumption that the biaxiality is small, as Chen and Noolandi 
previously suggested. This study showed indeed a small biaxiality, being 
thus consistent with its hypothesis. The effect of using a finer 
spatial grid was that an even smaller interfacial tension value $\gin
LD/k_BT=0.156\pm0.001$ was found (which agrees with recent simulational data
obtained by Vink and coworkers \cite{vink.schilling:2005, vink.wolfsheimer.schilling:2005}).  
The density and nematic order parameter profiles were
monotonic as Chen and Noolandi had also observed.
Only a small non-monotonicity  appeared in the density at the isotropic 
side in the case of normal
anchoring, i.e.~when the nematic director is normal to the plane
of the interface\cite{chen.noolandi:1992, chen:1993,shundyak.roij:2001}.   

Allen and coworkers have performed several simulation 
studies of closely related systems: they compared Onsager DFT with
simulations of soft rods \cite{al-barwani.allen:2000}
and with simulations of hard ellipsoids confined between walls 
\cite{allen:2000*a}; further they computed the
interfacial tension of the IN-interface in hard and soft ellipsoids from the
pressure tensor profile \cite{mcdonald:2000}, and studied 
capillary waves on the IN-interface of repulsive 
ellipsoids \cite{akino.schmid.ea:2001}. In these studies they found that
the interfacial tension agrees well with Onsager DFT, while the 
absolute coexistence densities are overestimated by theory.

In another simulation study nematic wetting films between
a hard wall and an isotropic fluid of hard rods have been observed \cite{rene}.
So far, however, the recent mean field predictions on the structure and
tension of the IN-interface in long cylindrical rods have not been tested
against computer simulations. 

If one sets out to compare mean-field results to computer simulations, the
role of fluctuations needs to be considered. In the following section we
briefly summarize the concept of capillary waves, and in section 
\ref{intrinsic} we discuss how we compare mean-field ``intrinsic'' profiles
to our simulation results.

\subsection{Capillary waves}
At finite temperature, an interface is never planar because of 
thermal fluctuations (``capillary waves''). 
As each deviation from the planar shape costs 
energy, the capillary wave spectrum is determined by the balance
between thermal energy and interfacial tension (and other
contributions to the free energy cost of the interface, such as bending 
rigidity.)
In the simplest case, the capillary wave spectrum is goverend by the free
energy cost of the interfacial area, which is added due to the undulations.
Assuming that the interface position can be parametrized by a function
$h(x,y)$, and that the local distortions are small, the wave spectrum
is predicted to be
\begin{equation}
\label{eq:capwaves}
\langle |h({\bf q})|^2\rangle = \frac{k_BT}{\gamma q^2} \quad ,
\end{equation}
where $h({\bf q})$ is the Fourier transform of
the local position $h(x,y)$ of the fluctuating surface,  
${\bf q}$ is the wavevector and
$\gamma$ is the interfacial tension. Note that this expression cannot hold
for $q\to 0$, because the amplitude would diverge. In real systems, gravity or
finite system size, for example, introduce a cutoff at small $q$. 
For an introduction to the subject of capillary waves see for example 
the book by Safran\cite{safran:1994}.
  
However, the capillary wave spectrum of the IN-interface cannot be described
in this simple manner. Li\-quid
crystals exhibit long-range elastic interactions, which keep the IN 
interface anisotropic up to very large length scales. (The interface becomes 
isotropic only on scales larger than the length over which deformation of 
the director field costs less energy than $\kb T$.) Also, the interfacial
tension $\gin$ is not a constant, but a function of the angle $\phi$ 
between the director and the interface. Therefore, 
fluctuations parallel to the director will produce different free energy 
costs than fluctuations perpendicular to the director. 
This separates the IN-interface 
from most other interfaces for which capillary waves have been studied.

Recently there have been several theoretical studies on waves on the 
IN-interface \cite{popanita.sluckin:2002, popanita.oswald:2003, 
elgeti.schmid:2005}. In particular, the IN-capillary wave spectrum 
has been analyzed by Elgeti and Schmid within Frank elastic theory and 
Landau-de Gennes theory \cite{elgeti.schmid:2005}. For the 
Landau-de Gennes approach their study shows that
\[ 
\frac{\kb T/\gamma^\parallel_{\rm IN}}{\langle |h({\bf q})|^2\rangle} = q^2 + 
\hat{q}^2_y\left[b q^3 + {\cal O}(q^4)\right] \quad ,
\]
where $\gamma^\parallel_{\rm IN}$ is the interfacial tension for in-plane
alignment (i.~e.~$\phi = 0$), 
$\hat{q}_y$ is the component of the unit vector ${\bf q}/q$ in the
direction of the bulk director 
and $b$ is a constant, which contains information on 
the elasticity of the director field and the anisotropy of the 
interfacial tension. (A more detailed expression can be found in the 
article by Elgeti and Schmid\cite{elgeti.schmid:2005}.)
Thus, the spectrum is isotropic in the leading $q^2$--term, but shows a
strong anisotropy in the next order $q^3$--term. An anisotropic contribution
to the $q^2$--term, which one would expect to find because of the anisotropic
interfacial tension, is removed by the elastic interactions.
Both the theoretical work
and the computer simulation on repulsive ellipsoids by Akino, Schmid and Allen
\cite{akino.schmid.ea:2001} show that the amplitudes of
the waves are largest in the direction perpendicular to the director.
Here, we compare these results to the hard-spherocylinder case. 

\subsection{Comparing DFT and simulation}
\label{intrinsic}
The IN-transition in rods is of first order and it increases in strength with
increasing aspect ratio. For large aspect ratios the only relevant 
fluctuations in a finite simulation-box are therefore capillary waves.
For the further discussion, we assume that there is an ``intrinsic'' 
mean-field interface, which is broadened by capillary waves in a manner 
that is statistically uncorrelated to the properties of the ``intrinsic'' 
interface. This concept has been under debate, because it clearly fails 
for short wavelengths i. e. wavelengths of the order of the size of the
particles \cite{tarazona.chacon:2004, stecki:1998, napiorkowski.dietrich:1993,
  mecke.dietrich:1999, robledo.varea.ea:1991}. However, it has proven to be
successful in studies for which the short wavelength limit was not
relevant (see for example \cite{werner.ea:1997, werner.ea:1999,
  vink.horbach.binder:2005}). Here, we are interested in a comparison
of simulations to Onsager theory and Landau-de Gennes theory, {\it i.~e.}~we 
are not studying molecular details and therefore we assume the intrinsic 
profile concept to be valid.

We test which effect fluctuations have on our DFT results, 
by convoluting the ``intrinsic'' 
DFT profile $\rho^{\rm DFT}(z)$ with a simple Ansatz for the capillary 
wave spectrum
\begin{equation}
\label{Ph}
P(h) = \frac{1}{\sqrt{2\pi s^2}}\exp\left(-\frac{h^2}{2s^2}\right) \quad ,
\end{equation}
where $h$ is the local height of the interface, $s^2$ is the local mean-square
displacement of the interfacial height
\[
s^2 = \frac{1}{2\pi \gamma}\ln\left(\frac{q_{\rm max}}{q_{\rm min}}\right)
\quad ,
\]
$q_{\rm max}$ is the length of the maximum wavevector (given by the particle
diamter) and $q_{\rm min}$ is the length of the minimum wavevector (given by
the size of the box). The estimated apparent profile then becomes 
\[
\rho^{\rm est}(z) = \int^{\infty}_{-\infty}\, dh\, \rho^{\rm DFT}(z-h)P(h) \quad
.
\]
This Ansatz for the capillary wave spectrum certainly does not capture the
specific properties of the IN-interface described in the previous section. But
we will show in the results section, that it suffices to test whether a
quantity can be compared between DFT and simulation or not.

The article is structured as follows: First we introduce the model and
the simulation techniques. Then we discuss the interfacial profiles
and compare them to DFT results. In the third part we present 
the analysis of the capillary wave spectrum and in the last section 
we summarize.

\section{Model and order parameters}

\subsection{DFT}

The basic ingredient of density functional theory is the grand potential
$\Omega[\rho]$ given as a functional of the one-particle distribution
$\rho({\bf r},{\bf u})$, with
${\bf r}$ the center-of-mass of a rod and ${\bf u}$ the unit vector of
the long axis of a rod:
\begin{eqnarray}
\beta\Omega[\rho]&=&\int \D{\bf r} \D{\bf u}\rho({\bf r},{\bf
    u})\left(\log[\rho({\bf r},{\bf u})L^2D] -1-{\beta\mu}   \right)\nonumber\\
&-&\frac{1}{2}\int  \D{\bf r} \D{\bf u}   \D{\bf r'} \D{\bf
    u'}f({\bf r},{\bf u};{\bf r'},{\bf u'}  )\rho({\bf
    r},{\bf u})\rho({\bf r'},{\bf u'})\nonumber,
\end{eqnarray}
where $\beta=1/k_BT$, and $f$ is the Mayer function which equals $-1$ if
the rods overlap and $0$ otherwise \cite{onsager:1949}.
Here we focus on long hard spherocylinders with
a length $L$ much larger than the diameter $D$. In the limit
$D/L\rightarrow 0$, in which we perform the DFT calculations, Onsager's
second virial functional is expected to be accurate
\cite{onsager:1949,chen:1993,shundyak.roij:2001,shundyak.roij:2003}.
For a given chemical potential $\mu$ the equilibrium density minimizes the
functional, and the minimum condition $\delta\Omega[\rho]/\delta\rho(
{\bf r},{\bf u} )=0$ gives rise to a nonlinear integral
equation for the equilibrium profile $\rho({\bf r},{\bf u})$:
\[
\log[\rho({\bf r},{\bf u})L^2D]-\int \D{\bf r'} \D{\bf
    u'}f({\bf r},{\bf u};{\bf r'},{\bf u'}  )\rho({\bf r'},{\bf u'})=\beta\mu.
\]
The minimum value of the functional, which is obtained after insertion of the
equilibrium profile into the functional, gives the equilibrium grand
potential $-pV$ for a bulk system of volume $V$  at pressure $p$,
and $-pV+\gamma A$ for a system with a planar interfacial area $A$ (with
$\gamma$ the interfacial tension).
In the latter geometry we assume that the distribution is independent of
the in-plane coordinates $x$ and $y$, such that the Euler-Lagrange
equation is to be solved for
$\rho(z,\theta,\varphi)$, with $z$ the cartesian coordinate
perpendicular to the interface and $\theta$ and $\varphi$ the polar and
azimuthal angle of ${\bf u}$ with respect to the nematic director far
from the interface, respectively. Here we impose boundary conditions
such that the coexisting isotropic and nematic bulk phases are obtained
at $|z|\rightarrow\infty$. From the equilibrium profile one extracts
\cite{chen.noolandi:1992,chen:1993,shundyak.roij:2001} the
total density and the order parameter profile as 
$\rho(z)=\int d{\bf u}\,\rho(z,{\bf u})$ and
$S(z)=\int d{\bf u}  \, P_2({\bf n}\cdot{{\bf u}})  \rho(z,{\bf u})/\rho(z)$,
respectively, with $P_2(x)=(3x^2-1)/2$ the second Legendre polynomial.
The biaxiality profile is given by $\alpha(z)=(3/2)\int d{\bf
  u}\,\rho(z,{\bf u}) \sin^2\theta\cos(2\varphi)/\rho(z)$. We present
the numerical results of the corresponding profiles in
Figs. \ref{profile}, \ref{biax}, and \ref{profileT}. 
An accurate
numerical method \cite{shundyak.roij:2001} gives for the surface
tension  $\gamma_{||}^{DFT} LD/k_BT=0.156\pm0.001$ at parallel anchoring,
and   $\gamma_{\perp}^{DFT} LD/k_BT=0.265\pm0.001$ at normal anchoring.

\subsection{Simulations}
As in the Onsager model, we consider spherocylinders, 
each consisting of a cylinder of length $L$
and diameter $D$ capped by two hemispheres of diameter $D$. 
The interaction between two particles $i$
and $j$ is given by a pair potential of the form
\begin{eqnarray*}
\label{eq:pot}
  v_{ij} (r) &=&
  \begin{cases}
  \infty & r < D, \\
  0 & {\rm otherwise},
  \end{cases}
\end{eqnarray*}
where $r$ is the distance between the cylinder axes. 

To investigate the IN transition, the density and the average rod alignment are
used as order parameters. We denote the density as the dimensionless
quantity $\rho^\star=(\pi/4 )\, L^2D\, N/V$ where $N$ is the number of 
particles, and $V$ is the volume of the simulation box.
The average alignment is defined in terms of the orientational traceless 
tensor $\bf{Q}$  with the elements
\[
\label{eq:s2}
 Q_{\alpha\beta} = \frac{1}{2 N} \sum_{i=1}^N
   \left( 3 u_{i\alpha} u_{i\beta} - \delta_{\alpha\beta} \right),
\]
where $u_{i\alpha}$ is the $\alpha$ component ($\alpha = x,y,z$)  of the 
unit vector along the axis of particle $i$
 and $\delta_{\alpha\beta}$ the Kronecker delta.
Diagonalization yields three eigenvalues, 
$$
S,  \quad -(S-\alpha)/2,\quad  -(S+\alpha)/2,\quad
$$
where $S$ is the nematic order parameter and
$\alpha$ is the biaxiality, which is equivalent to the definition
used in the DFT calculations.

\section{Simulation method}

Simulations were performed in the canonical ensemble using an elongated 
box with periodic boundary conditions and dimensions
$\lx\times\ly\times\lz = 10\times 10\times 20 \, L^3$.  
By rotational and translational Monte-Carlo
moves phase space was accessed.
The aspect ratio of the rods was $\frac{L}{D}=15$. The choice of L/D=15 is
motivated by the following requirements: on the one hand, the IN
transition becomes weaker with decreasing L/D. At short L/D
fluctuations in the bulk phases come into play, 
making the comparison with DFT very
difficult. On the other hand the CPU time required to detect an overlap
between two rods increases with L/D. Thus a compromise has to be made. 
In a previous study on ellipsoids, L/D=15 had been chosen 
\cite{akino.schmid.ea:2001}. We used the same to be able to compare our 
results to this study. 

The system was prepared in isotropic--nematic coexistence and the 
interfacial planes were located in the $(x,y)$--plane. Due
to the periodic boundary conditions at least two interfaces were required.
The overall density was chosen between the coexistence values, which
we had previously determined by grand canonical simulations 
\cite{vink.wolfsheimer.schilling:2005}. 
With this geometry about $10^5$ particles were required.
Parallel interfacial anchoring as well as normal anchoring was investigated.

Normal anchoring is metastable, therefore the director would
rotate towards a parallel orientation during a sufficiently long simulation. 
The time-scale necessary for such a rotation is given by the time it takes to
cross the free-energy barrier for the creation of a defect containing a 
reoriented region. As the
system is relatively stiff with respect to director field bending, we expect
this barrier to be rather high. During our simulation the
mean z-component of the director decreased by less than half a percent, 
while the fluctuations around this value were much larger than that. 
We have produced some data sets using a constraint on the orientation 
of the director, as it was introduced by Allen and coworkers
\cite{allen.warren.ea:1996}. However, the results did not differ from
the ones obtained in unconstraint simulations. 

After the equilibration run of about $10^7$ MC
sweeps (i.~e.~moves per particle) the values of the density and 
the orientational order parameter far away from the 
interface did not change anymore. The bulk values were reached 
in clear plateaus and they were in agreement with the results 
from the grand canonical simulations. 
These conditions were used as evidence for an equilibrated state.
For data production about $1.5 \times 10^6$ MC sweeps were performed.

\section{Results}
\label{sec:res}

\subsection{Interfacial profiles}
\label{sec:prof}

\begin{figure}
\begin{center}
  \includegraphics[clip,width=8cm]{Fig1}

\caption{\label{profile} 
Density and order parameter profiles for parallel anchoring.\\
\underline{Upper graph:} Simulation results of the reduced density $\rho^{*}$
(triangles) and the nematic order parameter $S$ (squares).
Dashed and solid lines are tanh-fits, which serve to guide the eye.
The bulk values of the density according to simulations in the grand-canonical
ensemble \cite{vink.wolfsheimer.schilling:2005} are indicated by a dotted
line. 
The shift $\delta$ between the inflection points of the profiles
comes out clearly. \underline{Inset:} Decay of the density and the order 
parameter to their isotropic bulk values on a log scale. Both quantities 
have the same decay length.\\
\underline{Lower graph:} Results from DFT calculations for density
(dashed line) and nematic order prameter (solid line) profiles.\\
Note that Onsager DFT overestimates the coexistence densities, while the 
shift $\delta$ between the interfaces agrees well. Differences between DFT and
simulations due to fluctuations are discussed in the text.
}
\end{center}
\end{figure}

\begin{figure}
\begin{center}
\includegraphics[clip,width=8cm]{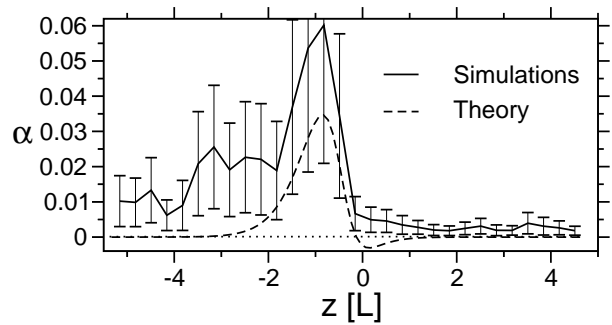}

\caption{\label{biax} 
Biaxiality profile from simulations (solid line)
and DFT calculations (dashed line) for parallel anchoring.
}
\end{center}
\end{figure}

\begin{figure}
\begin{center}
\includegraphics[clip,width=8cm]{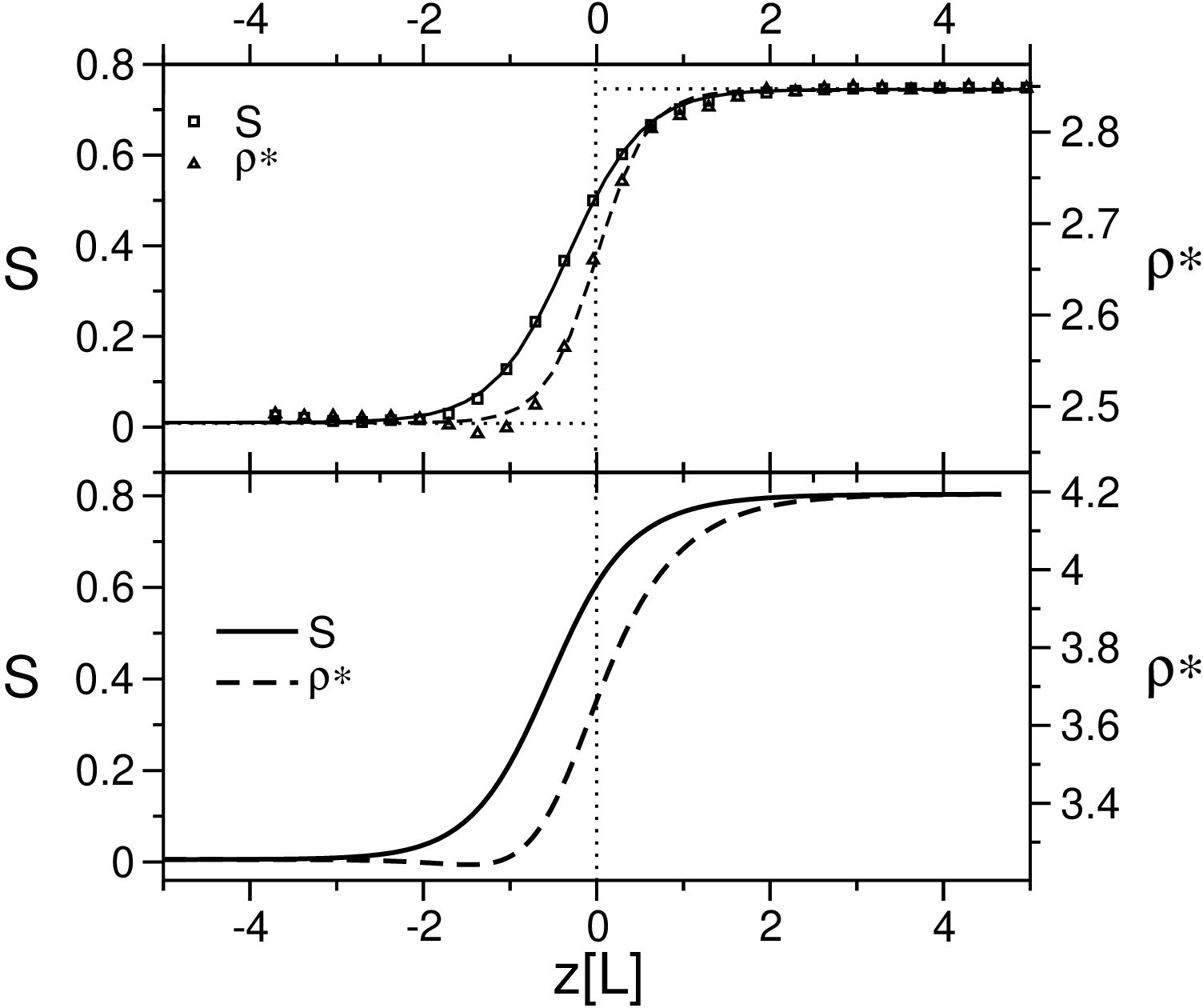}

\caption{\label{profileT} 
Density and order parameter profiles for 
normal anchoring. \\
\underline{Upper graph:} Reduced density $\rho^{*}$
(triangles) and nematic order parameter $S$ (squares).
Dashed and solid lines are tanh-fits, which serve to guide the eye.
The bulk values of the density according to simulations in the grand-canonical
ensemble \cite{vink.wolfsheimer.schilling:2005} are indicated by a dotted
line.\\
\underline{Lower graph:} DFT calculations for density (dashed line)
and order parameter (solid line) profiles. Both in simulation and theory, 
the density profile shows a small non-monoticity on the isotropic side.
}
\end{center}
\end{figure}

%

To investigate the interfacial profiles, the box was devided into $60$
slices in $z-$direction, and the density, the nematic order parameter,
and the biaxiality order parameter were sampled in each slice.
By averaging over $3 \times 10^5$ profiles the statistical error on
the density, which is much larger 
than the error on the order-parameter, became reasonably small and a clear 
biaxiality signal could be obtained.

Note, that the total average volume of nematic phase is
conserved, but that the center of the nematic layer
can shift along the $z-$direction without creating any energetic or
entropic  costs. 
To account for the freedom of the system to shift in the $z-$direction, 
 the profiles had to be centered before sampling. We shifted them such
 that their centers of mass lay in the middle of the simulation box.
(Fluctuations of the total amount of 
nematic phase lead to broadening of
the averaged profiles \cite{tepper.briels:2002, binder:1982}. 
This effect will be discussed in section \ref{sec:blocking}.)
The resulting profiles for parallel anchoring from theory as well as 
simulation are plotted in Figs.~\ref{profile} and \ref{biax}.  

Figs.~\ref{profile} and \ref{biax} show that the interface induces both 
nematic order and biaxiality on the isotropic side near the interface. 
This is in agreement with DFT calculations. 
The absolute values of the densities, however, differ considerably. 
This effect has been observed in similar systems by Allen and 
coworkers \cite{allen:2000*a, al-barwani.allen:2000}.
The induced nematic order on the isotropic side is reflected in the
shift between the inflection points of the 
density profile and the order parameter profile, 
$\delta=0.38(0)\, L$. DFT gives a similar result, 
$\delta=0.45\,L$ \cite{shundyak:diss}. 
Also in computer simulations of soft spherocylinders  
$\delta=0.37(4)\,L$ was found\cite{al-barwani.allen:2000}. 
For macroscopic system sizes, this quantity is not a constant because it is
affected by capillary waves. However, the profiles are almost symmetric with
respect to their inflection points. Therefore convolution with a symmetric
capillary wave spectrum \Eq{Ph} will hardly shift the inflection points and
thus will not lead to changes in $\delta$ unless very large systems 
are considered. We tested this
both, by changing the size of the simulation box and by computing the apparent
profiles $\rho^{\rm est}(z)$ and $S^{\rm est}(z)$ from the DFT profiles. The
changes in $\delta$ were of the order of $10\%$.

DFT also predicts that the density and order parameter
profiles should decay on the same length-scale, because the 
correlation length $\xi$ is the only characteristic 
length scale for these decays. The inset of \fig{profile} shows simulation
results for both profiles on a logarithmic scale revealing agreement with the
theoretical prediction. (For the system size used in the simulation, this
property is not affected by capillary waves, because it is measured 
far away from the interface.)

\Fig{profileT} shows the profiles for normal anchoring. Again there is good
agreement between simulation and theory. In particular, both show a weakly
non-monotonic behaviour of the density profile on the isotropic side. 
For increasing system-sizes this dip is smoothened out by interfacial
broadening. 

The present simulations confirm the theoretical claims \cite{chen:1993,
  shundyak.roij:2001, shundyak.roij:2003} that,
for the homeotropic (biaxial) IN-interface, the density and order parameter
profiles are monotonic and that the interfacial biaxiality is small. 
For the uniaxially symmetric
IN-interface the simulations confirm the theoretically predicted small
non-monotonic feature in the density profile. 

\subsection{Blocking analysis}
\label{sec:blocking}

In order to analyze capillary waves the interfacial position has to be
extracted from the simulation data. This requires the choice of a definition
of what constitutes the interface. Different choices of definition can 
lead to differences in the capillary wave spectra, in particular for 
short wavelengths. The simplest
approach is to cut the system into blocks of length $L_z$ and cross section
area $B\times B$ and to compute the Gibbs dividing surface for each block. 
However, the smaller the block, the larger are the fluctuations of the 
densities of the coexisting phases around their bulk values. This problem is
circumvented by the method we describe below and similar methods
\cite{akino.schmid.ea:2001, vink.horbach.binder:2005} which are based on 
sampling the differences between local profiles for each block in a 
configuration and the overall profile of the same configuration. 
These methods do not take into
account the molecular details of the interface and are therefore not useful
for studies of the capillary wave spectrum at short length-scales.
If one is interested in these, the pivot particle method by 
Tarazona and Chac{\'o}n \cite{tarazona.chacon:2004} can be used, which 
proceeds via a different route: First 
particles, which form the outmost layer of one phase, are identified. Then the
minimal surface through the positions of these particles is constructed.
Iteratively, more particles from the same phase which lie close to this 
surface are added and a new minimal surface is constructed until all surface
particles are incorporated. This method has shown to produce considerable
differences in capillary wave spectra at short wavelengths, and it would be
worthwhile applying it to the IN-interface. However, as we are
interested in a comparison with Landau-de Gennes theory, a continuum theory, 
we use a method that is less accurate on the molecular level. 

\begin{figure}
\begin{center}
\includegraphics[clip,width=7.5cm]{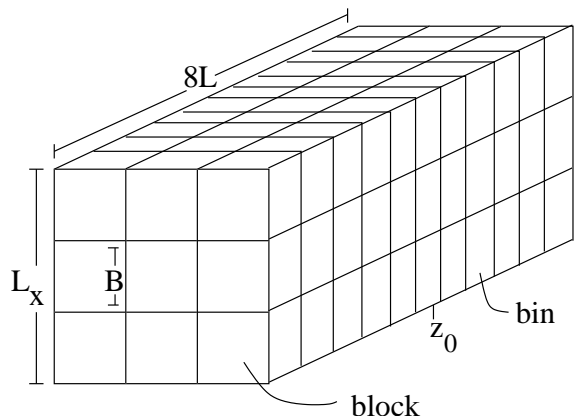}

\caption{\label{sketchBlocking} 
  Sketch of the blocking scheme. Details can be found in the text.
}
\end{center}
\end{figure}

\begin{figure}
\begin{center}
\includegraphics[clip,width=7.5cm]{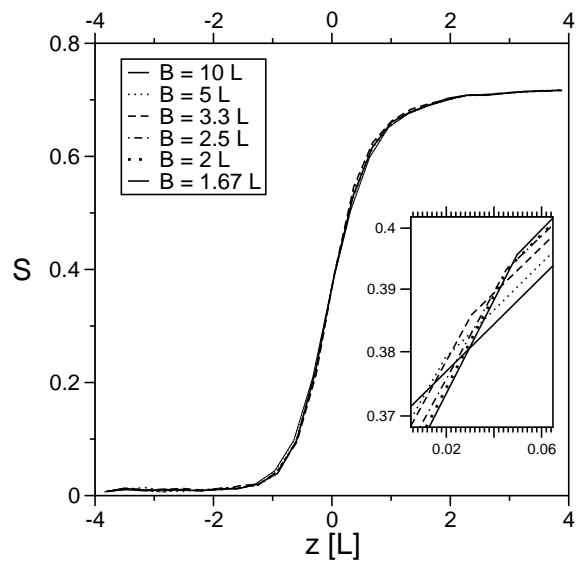}

\caption{\label{block_prof_L} 
  Interfacial profiles for various block sizes. 
The inset shows a close-up view with a logarithmic 
  scale of the profile near the center. 
}
\end{center}
\end{figure}

\begin{figure}
\begin{center}
\includegraphics[clip,width=7.5cm]{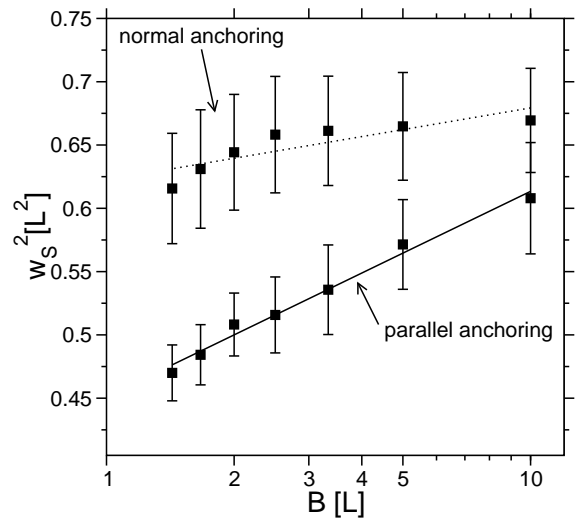}
\caption{\label{width2_LT} 
  Squared interfacial width $w_S^2$ versus $\ln(B)$ for
  parallel anchoring (solid line) and normal anchoring (dotted line). 
}
\end{center}
\end{figure}

In order to analyze the capillary waves, we used the following procedure:
We first determined the average plane of the interface (averaged over the
whole box cross section) and then measured undulations around this reference 
plane. It is, however, not sensible to determine a reference
plane for each individual system snapshot, as the position of the averaged 
profile is itself a fluctuating quantity. 
Instead we used averages over $10^4$ MC sweeps
to determine stable reference planes. 
For each snapshot, we shifted the system such that its center of mass 
was in the center of the box. Then the box was divided
into two parts, each containing one of the two interfaces. 
The interfaces were then investigated separately. 
In order to determine the reference plane of each interface we 
fitted the data with
\begin{equation}
  \label{eq:tanhs}
  S\left(z\right)=\frac{1}{2}\left(S_{N}+S_{I}\right)+
  \frac{1}{2}\left(S_{N}-S_{I}\right)\,
  \tanh\left(\frac{z-z_{0}}{w_{S}}\right)
\end{equation}
where $S_{N}$ and  $S_{I}$ are the bulk nematic and isotropic order parameters
respectively. 
The width $w_S$ and the position of the point of inflection $z_0$ are fit
parameters. Obviously, the profiles will not be perfect hyperbolic tangent
functions, because of the asymmetry in orientational order between the two
phases. However, the fit only serves to determine an average interfacial 
position as a basis for the following analysis: 
As sketched in \fig{sketchBlocking}, the box was then devided into 
square blocks in the $(x,y)$-plane with side lengths $B$ varing from $L_x / 7
$ to $L_x$. (Note that even the smallest block has a side length, which is
larger than a particle length. Thus we are far away from the molecular diameter
$D$.) Each block was then divided into  $25$ bins 
in the interval $\left[ z_0 - 4\,L,\; z_0 + 4\,L\right]$ in
$z$-direction. 
The profiles which we measured in these subsystems were again 
fitted by the tanh-function \eq{eq:tanhs} in order to find their points of 
inflection. All subsystem profiles for a given block size $B$ were centerend at
their points of inflection. Then {\bf Q} was averaged over all profiles 
and the average profiles were fitted by \eq{eq:tanhs} again in order to
sample the squared interfacial width $w_S^2$ and the interfacial positions 
$z\left(x_0,y_0\right)$
(where $x_0,y_0$ denotes the position of the block center).
$255$ system snapshots, {\it i.~e.~}$510$ snapshots of interfaces, have been used 
for the blocking analysis.
The smallest subsystem contained about $7500$ particles, 
which was sufficient to find a clear order parameter signal.

In order to make sure that the snapshots were decorrelated, we checked 
the autocorrelation functions of $w_S^2$ and $z$ for various
block sizes. 
We also ruled out possible effects due to the
finite bin size by comparing results for different discretisations. 
The size of a bin is determined by the following compromise: On the one hand 
a small bin size provides a good resolution of the
profile, but on the other hand the computation of order parameter average 
requires at least 100 particles.
For bin sizes between $10\,D$
and $20\,D$ the interfacial width as well as the computed interfacial tension
remained constant within the numerical accuracy.


\Fig{block_prof_L} shows the averaged order parameter profiles
for various block sizes. If a capillary wave spectrum is of the form of 
\eq{eq:capwaves} the interfacial width behaves 
as $w_S \propto \sqrt{\ln(B)}$ \cite{binder:1982}. 
\Fig{width2_LT} shows $w_S^2$ vs.~$\ln(B)$ for parallel and normal anchoring.
For parallel anchoring we find the predicted linear dependence. For
normal anchoring there might be deviations. However, the accuracy of the
simulations does not allow for a more detailed statement. 
The error-bars have been determined
by the ``jackknife-method'' \cite{efron:1979}.

\subsection{Analysis of the capillary wave spectrum}
\label{sec:capillary}

\begin{figure}
\begin{center}
\includegraphics[clip,width=7.5cm]{Fig7}
\caption{\label{fourier_L} 
Inverse mean-squared fourier components of the interface position
versus wave number squared ($m=q(\lx/2\pi)$) for parallel anchoring. 
}
\end{center}
\end{figure}

\begin{figure}
\begin{center}
\includegraphics[clip,width=7.5cm]{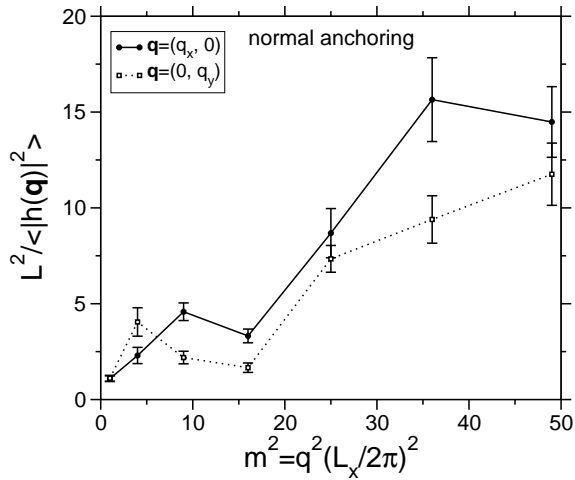}
\caption{\label{fourier_T} 
Inverse mean-squared fourier components of the interface position 
versus wave number squared for normal anchoring.
}
\end{center}
\end{figure}

\begin{figure}
\begin{center}
\includegraphics[clip,width=7.5cm]{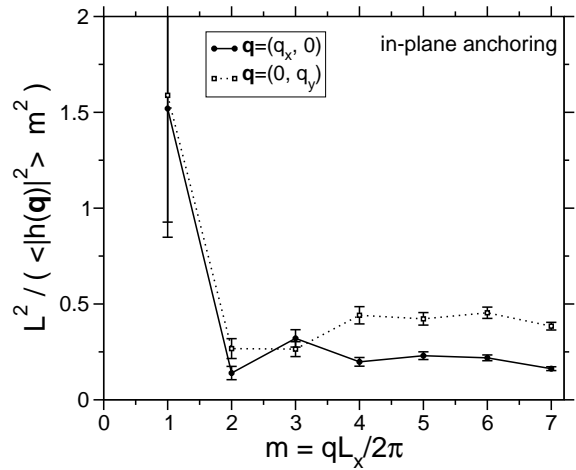}
\caption{\label{fourier_NL} 
$L^2/\left(\langle |h(\bf{q})|^2\rangle m^2\right)$. 
For $m>2$ amplitudes are proportional to $1/q^2$. The anisotropy of the
interfacial tension shows already on the $q^2$-level.
}
\end{center}
\end{figure}
 
A similar analysis as described in the previous section was used in
order to measure the capillary wave spectrum. For blocks of size
$L_x/14 \times L_y$, 
we sampled the distances between the points of inflection of the local 
profiles and the point of inflection of the reference profile. 
In order to test the stability of our
results, we also applied the method described by Akino and 
coworkers\cite{akino.schmid.ea:2001}. 
The accuracy was less good, but the results remained the same.
Further we checked that the distribution of interfacial heights $P(h)$ was 
Gaussian, indicating that the system was properly thermalized. 
Correlations between the undulations decayed within $10^4$ MC sweeps for the
larger values of $q$ and $5 \cdot 10^4$ for the smallest value. Hence we are
confident that the snapshots over which we averaged were sufficiently 
decorrelated. 

Figs.~\ref{fourier_L} and \ref{fourier_T} show 
$L^2/\langle |h(\bf{q})|^2\rangle$ vs. wave number squared
for parallel and normal anchoring.
(We have plotted ${\bf{m} }= {\bf{q}}(\lx/2\pi)$ 
instead of $q$ to facilitate direct comparison to
the results of Akino and co-authors \cite{akino.schmid.ea:2001}.)
Again the error-bars
have been determined by the ``jackknife-method''.
In the case of parallel anchoring the director was oriented along the
$y$-axis. 
As expected, the spectrum is anisotropic for parallel anchoring and 
isotropic for normal anchoring.
For parallel anchoring, we find in 
agreement with theory \cite{elgeti.schmid:2005} that fluctuations
parallel to the director (dotted line) are smaller than fluctuations
perpendicular to the director (solid line). 

The magnitude of the fluctuations compares well with the results Akino
and co-authors have obtained for ellipsoids \cite{akino.schmid.ea:2001}. 
However, their results deviate from the
$q^2$ dependence already for $m^2 \approx 10$.
To check whether there are contributions of higher order than $q^2$, we 
have plotted $L^2/(\langle |h({\bf{q}})|^2\rangle m^2)$ versus wave
number for parallel anchoring in \fig{fourier_NL}. 
For $m>2$, both components are almost constant.
(The large values for very small wave numbers are 
due to the finite system size.)
Parallel to the director, there might be small deviations from the $q^2$ 
dependence, but the results are not accurate enough to allow for 
detailed conclusions.   
Additionally we find that there is a difference 
in the absolute value $L^2/\left(\langle |h({\bf{q}})|^2\rangle m^2\right)$ 
between parallel and perpendicular contributions. This means that for the
regime of wave-numbers studied in this work, the
anisotropy of the interfacial tension does enter already on the $q^2$-level.

\begin{figure}
\begin{center}
\includegraphics[clip,width=7.5cm]{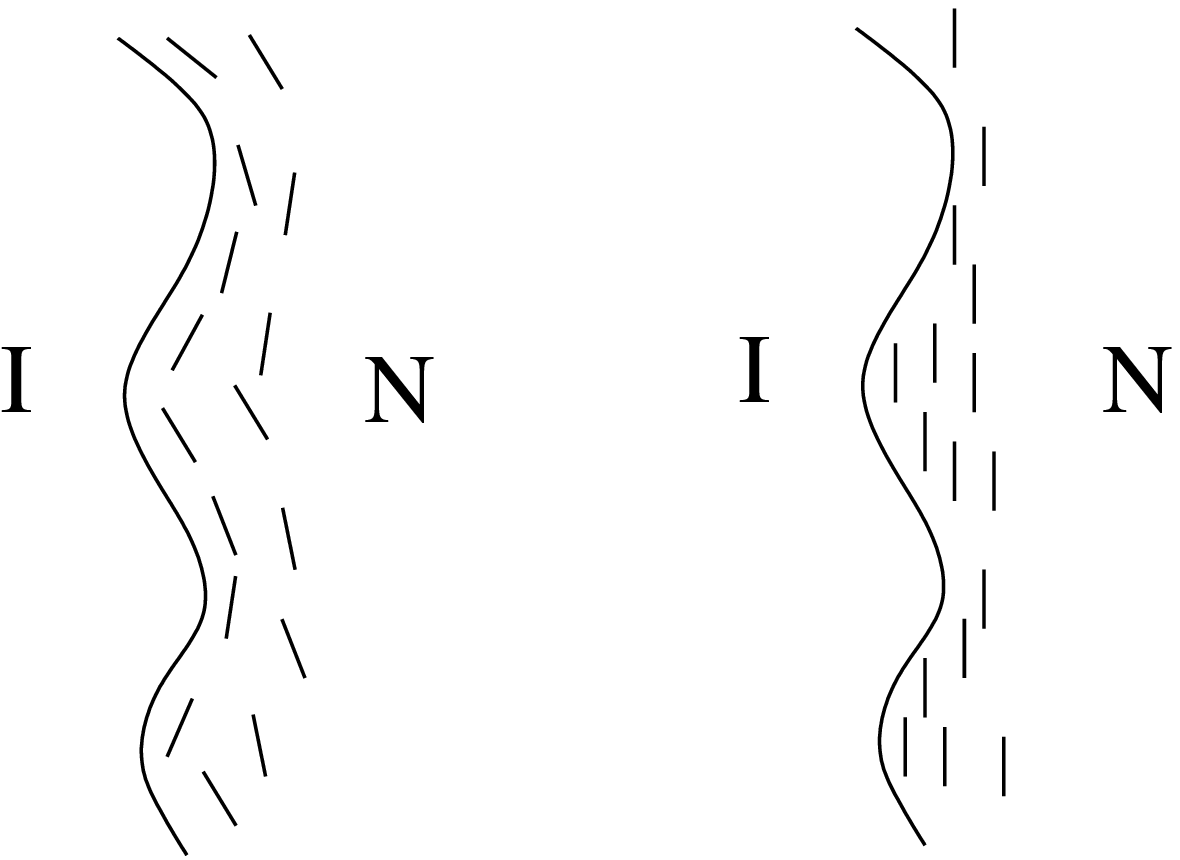}
\caption{\label{bendaniso} 
Sketch of capillary waves:\\ 
\underline{Left:} Strong interfacial anchoring $\longrightarrow$ 
anisotropy of the spectrum is dominated by elasticity\\
\underline{Right:} Strong bending rigidity $\longrightarrow$ anisotropy of 
the spectrum is dominated by the anisotropy of the interfacial tension 
}
\end{center}
\end{figure}

\Fig{bendaniso} shows a sketch of capillary waves in two extreme cases:
in the case on the left, parallel anchoring is so strong that the particles'
orientations have to follow the shape of the interface. The anisotropy of the 
spectrum is then dominated by contributions from bending elasticity. 
In the case on the right, the bending
rigidity is so strong, that the anisotropy of the spectrum is dominated 
by contributions from 
the anisotropy of the interfacial tension. From the fact that we do not 
observe contributions due to bending, but
strong contributions due to interfacial tension anisotropy, we conclude that
spherocylinders fall into the category sketched on the right.

\section{Discussion and summary}

We have presented a comparative study of computer simulations and Onsager DFT
for the isotropic-nematic interface in hard spherocylinders. Although Onsager
theory overestimates the isotropic-nematic coexistence densities, interfacial 
properties such as the the shift between the density 
profile and the nematic order parameter profile are in agreement. 
This observation accords with the studies 
by Allen and coworkers on ellipsoids \cite{allen:2000*a} and soft 
spherocylinders \cite{al-barwani.allen:2000}.  

Our simulations confirm that, for the case of in-plane orientation of the
director at the interface, the density and order parameter
profiles are monotonic and that the interfacial biaxiality is small. 
For the case of normal orientation the simulations agree with the 
theoretically predicted small non-monotonic feature in the density profile. 

The capillary wave spectrum of the interface (in the case of parallel 
anchoring) is anisotropic, as expected. 
In contrast to the case of ellipsoids, we
find that $L^2/\langle |h({\bf{q}})|^2\rangle \propto q^2$ for the entire
range of wave vectors which could be accessed in this study. Also we observe
anisotropic contributions already on the $q^2$-level. We interpet both facts
as signals of the a very stiff director field, which prohibits bending along
the fluctuating interface. 

\acknowledgments

We thank the Deutsche Forschungsgemeinschaft (DFG) for support
(TR6/D5) and K.~Binder, R.~Vink, P.~van der Schoot and J.~Horbach
for helpful suggestions. TS is supported by the Emmy Noether program of the
DFG and by the MWFZ. 
Allocation of computer time on the JUMP at the Forschungszentrum
J\"ulich is gratefully acknowledged.
This work is part of the research program
of the ``Stichting voor Fundamenteel Onderzoek der Materie
(FOM)'', which is financially supported by the ``Nederlandse
organisatie voor Wetenschappelijk Onderzoek (NWO)''.

\bibstyle{revtex}
\bibliography{man}

\end{document}